\documentclass[twocolumn,showpacs,preprintnumbers,amsmath,amssymb]{revtex4}

\usepackage{graphicx}
\usepackage{dcolumn}
\usepackage{bm}

\begin{document}


\title{Density-Wave and Antiferromagnetic States of Fermionic Atoms in Optical Lattices}

\author{Takuji Higashiyama}
\author{Kensuke Inaba}%
\affiliation{Department of Applied Physics, Osaka University, Suita, Osaka 565-0871, Japan}

\author{Sei-ichiro Suga}
\affiliation{Department of Applied Physics, Osaka University, Suita, Osaka 565-0871, Japan}

\date{\today}

\begin{abstract}
We study the two-band effects on ultracold fermionic atoms in optical lattices by means of dynamical mean-field theory. 
We find that at half-filling the atomic-density-wave (ADW) state emerges owing to the two-band effects in the attractive interaction region, while the antiferromagnetic state appears in the repulsive interaction region. As the orbital splitting is increased, the quantum phase transitions from the ADW state to the superfluid state and from the antiferromagnetic state to the metallic state occur in respective regions. Systematically changing the orbital splitting and the interaction, we obtain the phase diagram at half-filling. The results are discussed using the effective boson model derived for the strong attractive interaction. 
\end{abstract}

\maketitle

\section{INTRODUCTION}
Ultracold fermionic atoms in optical lattices have attracted considerable attention. By loading fermionic atoms into optical lattices, diverse interaction configurations can be introduced. Furthermore, Feshbach resonances provide the means for controlling both the strength of the interaction between fermionic atoms and its sign \cite{Chen2005}. Fascinating quantum many-body phenomena have been revealed by the combination of these experimental techniques. 
The topological change in the Fermi surface of $^{40}{\rm K}$ fermionic atoms was observed by increasing the band-filling \cite{Kohl2005}. By controlling the atomic interaction, a band insulator in the lowest band was produced and then the partially populated higher bands were observed. 
The Mott insulating state for $^{40}{\rm K}$ was realized quite recently by adequately tuning the ratio between the interaction and kinetic energy \cite{Jordens2008}. Three features were identified; a suppression of doubly occupied lattice, a reduction of the compressibility, and the gapped mode in the excitation spectrum. It was argued that the results pave the way for future studies of spin ordering. 
For $^6{\rm Li}$ fermionic atoms with attractive interaction, a superfluidity of fermionic atom pairs was observed \cite{Chin2006}. By increasing the depth of the lattice potential near the Feshbach resonance, a superfluid-insulator transition was observed. 
In the experiments it was argued that the usual single-band model was no longer applicable, because the strength of the on-site interaction exceeded the gap between the lowest and the next-lowest bands. Accordingly, the effects of the higher bands have to be taken into account for detailed investigations.

A Mott insulating state \cite{Kollath2006,Zhai2007,Moon2007,Chien2008,Helmes2008,Higashi,Goldman} and/or ordered states \cite{Andersen,Wu,Koga2008,Snoek2008} for fermionic atoms in two- or three-dimensional optical lattices have been investigated by various theoretical methods. 
Recently, we have investigated the superfluid-insulator transition of fermionic atoms in optical lattices beyond the conventional mean-field approximation, taking two-band effects into account \cite{Higashi}. It was shown that the Mott insulating state appears for the adequately strong interaction region at half-filling. For the Mott insulator in the repulsive interaction the fermionic atoms in each site occupy both orbitals, while in the attractive interaction the bosonic fermion pairs occupy either of the two orbitals in each site. The results suggest that other ordered states may be possibly induced: the atomic-density-wave (ADW) state for the attractive region and the antiferromagnetic (AF) state for the repulsive region. However, an issue whether these ordered states compete or coexist with the superfluid state or Mott insulating states has not yet been well investigated for two-band lattice fermionic systems.

In this paper, we investigate the ordered state of ultracold fermionic atoms in three-dimensional optical lattices, taking the two-band effects into account. For this purpose, we make use of a dynamical mean-field theory (DMFT) \cite{Georges1996}, which enables us to treat local correlation effects precisely. We show that for the attractive interaction region the ADW state is stable for a small orbital splitting region and that the transition to the superfluid state takes place with increasing the orbital splitting. For the repulsive interaction region, the AF state appears in a small orbital splitting region. As the orbital splitting increases, the AF state changes to the metallic state. For large orbital splittings, the band insulating state comes into existence in both regions.

The paper is organized as follows. In Sec. II, we introduce the model Hamiltonian and explain the two-site DMFT method to investigate the ADW and AF states. 
In Sec. III, the numerical results for the attractive and repulsive regions at half-filling are shown. We discuss the most stable state for given parameters by comparing their energies. The results are summarized in the phase diagram. In Sec. IV, we discuss the numerical results in comparison with the effective boson model, which is derived for the strong attractive region. It is shown that both results are consistent with each other. A brief summary is given in Sec. V.

\section{MODEL AND METHOD}
Let us consider the fermionic atoms in a optical lattice potential: $V(\bold{r})=V_0 (\sin^2 kx + \sin^2 ky + \sin^2 kz)$. In the low-tunneling $V_0 \gg E_r$, where $E_r =\hbar^2 k^2/2m$ is the recoil energy, each lattice potential is regarded as a harmonic one \cite{Zwerger2003,Hofstetter2006}. We investigate the effects of the lowest and next-lowest orbitals, so that the three-fold degeneracy of the next-lowest orbitals is neglected for simplicity. 
The hopping integrals between the lowest orbitals $(t_1)$ and between the next-lowest orbitals $(t_2)$ satisfy the relation $t_2 \sim \sqrt{V_0/E_r} \, t_1$. Since $V_0/E_r \lesssim 10$ in the experiments \cite{Kohl2005,Chin2006}, we approximately set that $t_1=t_2 \equiv t$. 
The following interactions are considered: the on-site intraorbital interactions for the lowest orbital $(U_1)$ and for the next-lowest orbital $(U_2)$, the interorbital interaction $(U')$, and the interaction corresponding to the Hund coupling $(J)$. The coupling constants of these four interactions satisfy the relations $U_2=(3/4)U_1$ and $U'=J=U_1/2$. We set that $U_1=U_2 \equiv U$ approximately and thus $U'=J=U/2$.

The system is assumed to involve the same number of fermionic atoms in two different hyperfine states, which are described as the pseudospins. 
The simplified model Hamiltonian thus obtained reads 
\begin{eqnarray}
{\cal H} &=& \sum_{\langle i,j \rangle \alpha\sigma}(t_{}^{}-\mu\delta_{i,j})
             c_{i\alpha\sigma}^{\dagger}c_{j\alpha\sigma}^{} 
            + \frac{D}{2}\sum_{i\sigma}(n_{i2\sigma}-n_{i1\sigma}) \nonumber\\
         &+& U^{}\sum_{i\alpha} n_{i\alpha\uparrow}n_{i\alpha\downarrow} 
         + \sum_{i\sigma \sigma^{\prime}}
             (U^{\prime}-J \delta_{\sigma, \sigma'}) 
                n_{i1\sigma} n_{i2\sigma'}, 
\label{hami}
\end{eqnarray}
where $c_{i\alpha\sigma}$ is the fermionic annihilation operator for the state with pseudospin $\sigma$(=$\uparrow$,$\downarrow$) on orbital $\alpha$(=1, 2) in the $i$th lattice site, $n_{i\alpha\sigma}= c_{i\alpha\sigma}^{\dagger}c_{i\alpha\sigma}$, and the subscript $\langle i,j \rangle$ indicates the sum of the nearest-neighbor sites. $\mu$ is the chemical potential and $D$ is the splitting between the two orbitals. 
We assume that the intraorbital attractive interaction induces an $s$-wave superfluid state.

For the attractive region $(U<0)$ we examine the superfluid-insulator transition, turning our attention to the stability of the ADW state. For the repulsive region $(U>0)$ the phase transition between the metallic state and the insulating state is examined, laying stress on the stability of the AF state.

In DMFT, the lattice model is mapped onto a single impurity model connected dynamically to a heat bath. The Green's function is obtained via the self-consistent solution of this impurity problem. This retains nontrivial local quantum fluctuations missing in conventional mean-field theories. We apply here the two-site DMFT method \cite{Potthoff2001}, which allows us to study the Mott transitions of orbitally degenerate lattice fermions qualitatively \cite{Ono2003,Koga2004}. To study the superfluid of lattice fermions, we extend this method to the case when the superfluid order exists \cite{Higashi}.

In order to examine the ADW and AF states, we divide the bipartite lattice into two sublattices \cite{Georges1996,Chitra}. 
In this procedure, the local Green's function has the following form: 
\begin{widetext}
\begin{eqnarray}
{\hat{G}_{\alpha}^{}(\omega)} &=&
\int dz \rho(z) {\hat{G}_{\alpha}^{}(z,\omega)}, \\
{\hat{G}_{\alpha}^{-1}(z,\omega)} &=& 
 \left(
  \begin{array}{ccc}
   \omega+\mu-(-1)^{\alpha}\frac{D}{2}-\Sigma_{A,\alpha}(\omega) & -z \\
   -z & \omega+\mu-(-1)^{\alpha}\frac{D}{2}-\Sigma_{B,\alpha}(\omega)
  \end{array}
 \right),
\label{green}
\end{eqnarray}
\end{widetext}
where $\rho(z)$ is the density of states (DOS). $\Sigma_{A(B),\alpha}(\omega)$ is the self-energy of the orbital $\alpha$ for the $A (B)$ sublattice, which can be obtained by solving two effective impurity models.
We use a semicircular DOS, $\rho_{}(z)=4/(\pi W_{})\sqrt{1-4(z/W_{})^{2}}$, where $W$ is the band width. Since the hopping integral is assumed to be independent of $\alpha$, $W_{}=4t$ and the DOS are the same for both bands. The chemical potential is set to be $\mu=U/2+U'-J/2$ so that particle-hole symmetry can be satisfied. In this case, two bands by the orbitals $\alpha=1$ and $2$ together are half-filling. In the following, the hopping integral $t$ is used in units of energy.

\section{NUMERICAL RESULTS}
\begin{figure}[htb]
\begin{center}
\includegraphics[scale=0.65]{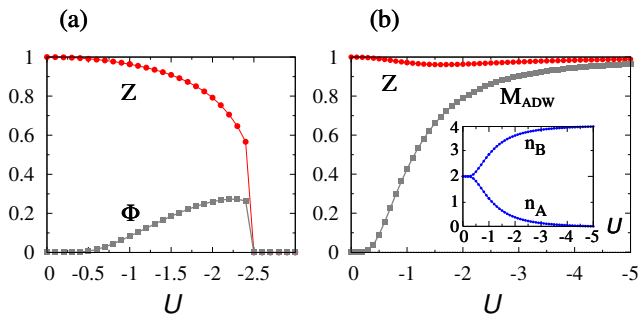}
\caption{(Color Online) (a) The quasiparticle weight $Z$ and the superfluid order parameter $\Phi$ as functions of $U(<0)$ for $D=0$. (b) The ADW order parameter $M_{\rm{ADW}}$ and $Z$ as functions of $U(<0)$ for $D=0$.
Inset: The atomic number per site of the $B(A)$ sublattice $n_{B(A)}=\sum_{\alpha, \sigma} \langle n_{B(A)\alpha\sigma} \rangle$. }
\label{fig1}
\end{center}
\end{figure}

For the attractive interaction region $U<0$, three types of states are considered: the superfluid state, the Mott insulating state, and the ADW state. 
To investigate these states, we calculate the superfluid order parameter $\Phi=\langle c_{i\alpha\downarrow}c_{i\alpha\uparrow} \rangle$, the quasiparticle weight $Z$, and the ADW order parameter $M_{\rm{ADW}}=(1/4) (n_{B}-n_{A})$, where $n_{B(A)}=\sum_{\alpha, \sigma} \langle n_{B(A)\alpha\sigma} \rangle$ with $n_{B(A)\alpha\sigma}=c_{B(A)i\alpha\sigma}^{\dagger}c_{B(A)\alpha\sigma}$ being the number operator of the $B(A)$ sublattice per site. $Z$ represents the coherent spectral weight of the Bogoliubov quasiparticle \cite{Garg2005}. Because of particle-hole symmetry, $\Phi$ is independent of $\alpha$.

We first calculate $Z$ and $\Phi$ for the investigation of the superfluid-Mott insulator transition, without dividing the system into two sublattices. 
In Fig. \ref{fig1}(a), the results for $D=0$ are shown. As $|U|$ increases, $Z$ decreases and jumps to $0$ with vanishing $\Phi$ simultaneously. The results indicate that the discontinuous quantum phase transition from the superfluid to the Mott insulator occurs. This Mott transition is caused by the two-band effects \cite{Higashi}. We find that the Bogoliubov quasiparticle is renormalized significantly towards the transition point.

We next calculate $M_{\rm{ADW}}$ and $Z$, dividing the system into two sublattices. 
The results are shown in Fig. \ref{fig1}(b). $Z$ is nearly equal to $1$ irrespective of $U$ and $M_{\rm{ADW}}$ increases monotonously towards 1 with $|U|$. The results indicate that for $D=0$ the ADW order is enhanced by the attractive interaction $U$. As shown in the inset of Fig. \ref{fig1}(b) the fillings of the neighboring sites approach 4 and 0, respectively, with increasing $|U|$. 
The results demonstrate that the imbalanced atomic numbers between both sublattices take place in the ADW state and for $|U| \gtrsim 4$ the almost fully-occupied and empty states emerge alternately.
Note that $n_{B}+n_{A}=4$ irrespective of $U$ nor $D$ due to particle-hole symmetry. 

\begin{figure}[htb]
\begin{center}
\includegraphics[scale=0.55]{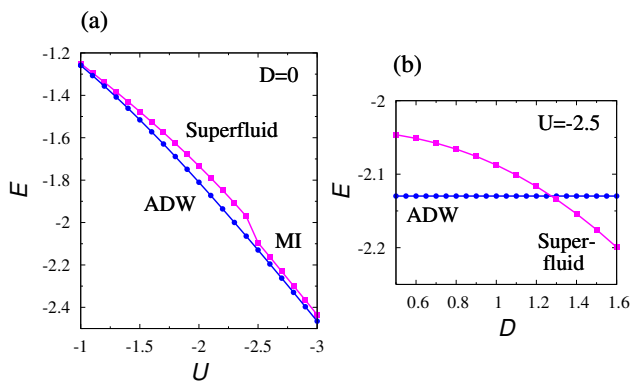}
\caption{(Color Online) The energies of three states as functions of (a) $U(<0)$ for $D=0$ (b)$D$ for $U=-2.5$. MI indicates the Mott insulator.}
\label{fig2}
\end{center}
\end{figure}

To determine the most stable state among these three states, their energies are compared. As in Fig. \ref{fig2}(a), we confirm that the ADW state is the most stable for all $U$ at $D=0$. 
For $|U|<1$ the energy of the ADW state is lower than that of the superfluid state, although their differences are so small in the present scale. 

We calculate $Z$, $\Phi$, and $M_{\rm{ADW}}$ for $D \neq 0$ by the same method. To investigate the most stable state among them, we compare the energies of these states. In Fig. \ref{fig2}(b), the results for $U=-2.5$ are shown as functions of $D$. As $D$ increases, the energy of the superfluid state decreases and crosses the energy of the ADW state. Accordingly, the discontinuous quantum phase transition to the superfluid state occurs at $D=1.27$.

\begin{figure}[htb]
\begin{center}
\includegraphics[scale=0.65]{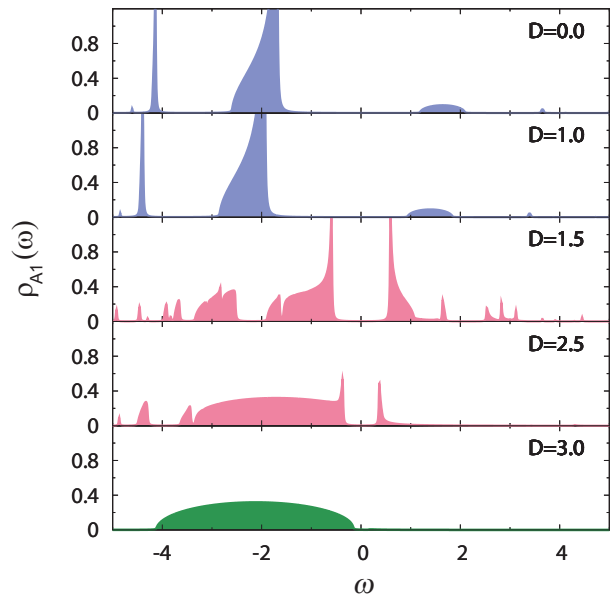}
\caption{(Color Online) The single-particle excitation spectra $\rho_{A1}(\omega)$ for several values of $D$ at $U=-2.5$.}
\label{DOS}
\end{center}
\end{figure}

We investigate the quantum phase transition from the spectral point of view. 
The single-particle excitation spectra (SPES) are defined as $\rho_{\gamma\alpha }(\omega)=-(1/\pi){\rm Im}{\hat{G}_{\alpha}(\omega+i\delta)} \: (\gamma=A,B)$, where $\delta$ is a small positive number, and $\rho_{B \alpha}(\omega)$ and $\rho_{A \alpha}(\omega)$ are given by the $(1,1)$ and $(2,2)$ components of the matrix ${\hat{G}_{\alpha}(\omega+i\delta)}$, respectively. 
Because of particle-hole symmetry, $\rho_{A1}(\omega)=\rho_{B2}(-\omega)$ and $\rho_{B1}(\omega)=\rho_{A2}(-\omega)$. 
For the superfluid and band-insulating states, the relation $\rho_{A\alpha}(\omega)=\rho_{B\alpha}(\omega)$ is satisfied. 
In Fig. \ref{DOS}, the SPES $\rho_{A1}(\omega)$ are shown for several values of $D$ at $U=-2.5$. 
The SPES for $D=1.0$ shifts to the low energy region as compared to that for $D=0$ with scarcely changing the spectral shape. Accordingly, the atomic numbers $n_A$ and $n_B$ for given $U$ are independent of $D$. 
These properties are characteristic of the ADW state. The spectral gap around $\omega=0$ is $\sim 2.3$, which is larger than $D$. The energy of the ADW state is thus independent of $D$ for the range shown in Fig. \ref{fig2}(b). 
For $D=1.5$, we find the incoherent spectral weights away from $\omega=0$, which indicate the significant renormalization of the Bogoliubov quasiparticle. As $D$ is increased, the enhanced incoherent spectral weights become inconspicuous and for $D=3.0$, the SPES exhibits a typical profile of the band insulator. 
These findings are consistent with the behavior of other quantities.

We have confirmed for the single-band attractive Hubbard model ($J=U'=D=0$) at half-filling that the superfluid-insulator transition never occurs as shown in the studies so far \cite{Garg2005,Toschi2005,Kyung2006}. At half-filling the superfluid state and the ADW state are degenerate, while in the filling deviated from the half-filling the superfluid state is the most stable \cite{Scalettar1989,Capone2002,Garg2005}. In the present system, the ADW state persists up to a certain value of $D$ for given $U$. This feature is caused by the two-band effects.

\begin{figure}[htb]
\begin{center}
\includegraphics[scale=0.65]{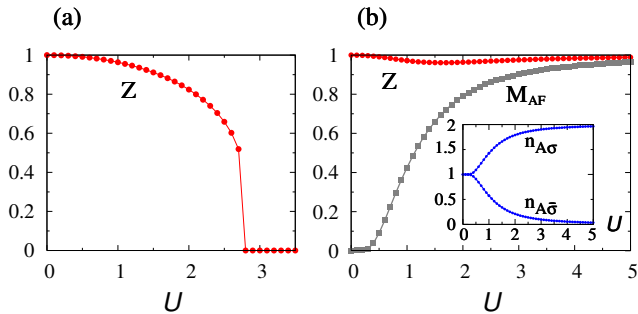}
\caption{(Color Online) (a) Quasiparticle weight $Z$ as a function of $U(>0)$ for $D=0$. (b) The ADW order parameter $M_{\rm{ADW}}$ and $Z$ at $D=0$. 
Inset: The atomic number per site of the $A$ sublattice with $\sigma$ ($\bar{\sigma}$) pseudospin $n_{A\sigma(\bar{\sigma})}=\sum_{\alpha} \langle n_{A\alpha\sigma(\bar{\sigma})} \rangle$. }
\label{D=0_r}
\end{center}
\end{figure}
\begin{figure}[htb]
\begin{center}
\includegraphics[scale=0.55]{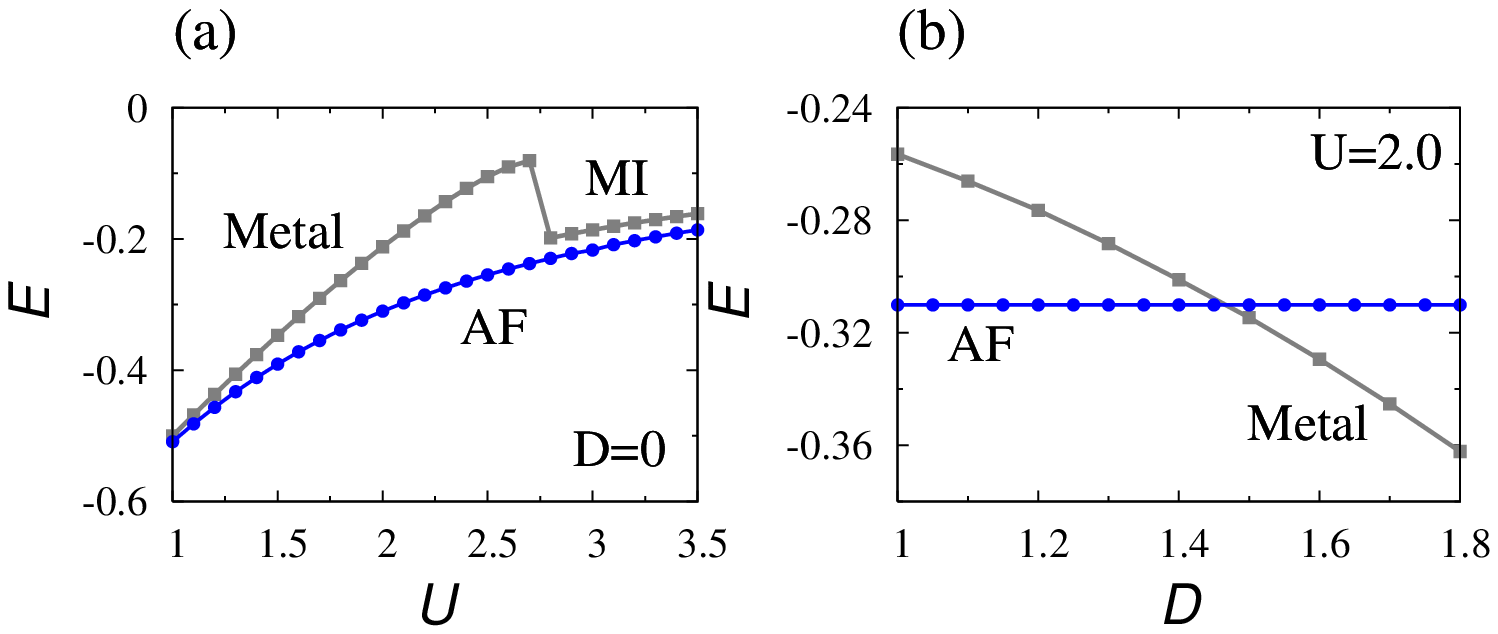}
\caption{(Color Online) The energies of three  states as functions of (a) $U(>0)$ for $D=0$ and (b)$D$ for $U=2.0$. MI indicates the Mott insulator.}
\label{Energy_r}
\end{center}
\end{figure}
For the repulsive interaction region $U>0$, we calculate $Z$ and the staggered magnetization of the sublattices $M_{\rm{stagg}}^{\gamma}=(1/2)(n_{\gamma\sigma}-n_{\gamma \bar{\sigma}})$ ($\gamma=A, B$) with $n_{\gamma\sigma}=\sum_{\alpha} \langle n_{\gamma\alpha\sigma} \rangle$ ($\gamma=A, B$) and $\bar{\sigma}$ being the opposite state of $\sigma$. 
The relations $M_{\rm{stagg}}^{A}=-M_{\rm{stagg}}^{B} \equiv M_{\rm{AF}}$ and $n_{A(B)\sigma}+n_{A(B) \bar{\sigma}}=2$ are satisfied due to particle-hole symmetry. 
We first calculate $Z$ for the investigation of the Mott transition, so that the system is not divided into two sublattices. As shown in  Fig. \ref{D=0_r}(a), the discontinuous transition from the metallic state to the Mott insulator occurs at $U=2.8$. 
We next calculate $M_{\rm{AF}}$ and $Z$, dividing the system into two sublattices. As shown in Fig. \ref{D=0_r}(b), $M_{\rm{AF}}$ increases toward $1$ and $Z$ takes the value close to $1$ with increasing $U$. The inset of Fig. \ref{D=0_r}(b) shows that the pseudospin state of the A sublattice approaches the fully-polarized one, as $|U|$ is increased. 
The behavior exhibits typical features of the AF state. 
We compare the energies among the Mott insulating state, AF state, and renormalized metallic state to determine the most stable state. 
As shown in Fig. \ref{Energy_r}, the AF state is the most stable for $D=0$ and the discontinuous phase transition to the metallic state occurs at $D=1.46$ for $U=2.0$. 
For $U<1$ the energy of the AF state is lower than that of the metallic state, although their differences are so small in the present scale.

\begin{figure}[htb]
\begin{center}
\includegraphics[scale=0.82]{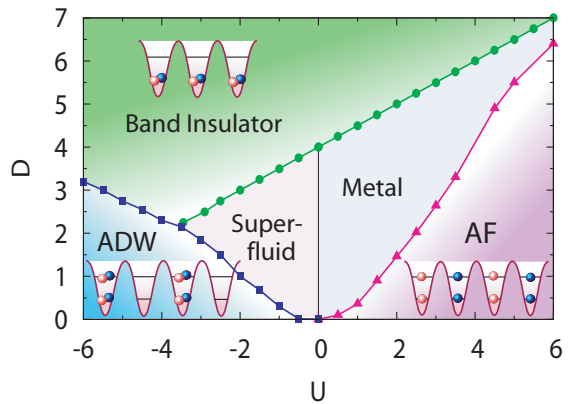}
\caption{(Color Online) Phase diagram for the two-band Hubbard model at half-filling. The spheres with blue (dark) color represent atoms with `up' pseudospin and the spheres with red (light) color represent atoms with `down' pseudospin.}
\label{PD}
\end{center}
\end{figure}

Systematically changing $D$ and $U$, we investigate the ground state of the two-band Hubbard model at half-filling. The results are summarized in the phase diagram shown in Fig. \ref{PD}. 
In the large $D$ region, the band insulator comes into existence \cite{Higashi}. In the small $D$ region, the ADW state and the AF state appear. As $D$ is increased for the attractive region the direct transition between the ADW state and the band insulating state is observed for $|U| \geq 3.5$, while for the repulsive region the metallic state appears between the AF and band insulating states even in $U=6$. 
It is considered that the features are caused by the difference in spin and orbital fluctuations. 
In the repulsive interaction region, spin and orbital fluctuations are enhanced as $U$ is increased \cite{Koga2002,Werner2007}, leading to the stabilized metallic state even for large $D$. 
In the attractive interaction region, on the other hand, spin fluctuations disappear because of the formation of a Cooper pair. In the ADW state, orbital fluctuations also vanish. The ADW state is thus considered to be robust, which yields the direct transition to the band insulating state with increasing $D$.

\section{DISCUSSIONS}
We investigate the stable ADW state using an effective boson model, which is appropriate for the strong attractive $U$ in the half-filling. 
To derive the effective boson model from the Hamiltonian (\ref{hami}), we first make a particle-hole transformation in the higher orbital. We then expand the transformed Hamiltonian in the strong attractive $U$ ($D \ll |U|$), where two fermions with different pseudospins make a hardcore boson. 
Within the fourth order expansion, the following effective model for the bosonic fermion pair can be derived, 
\begin{eqnarray}
{\cal H_{\rm{eff}}} &=& 
      t_{\rm{eff}}^{} \sum_{\langle i,j \rangle,\alpha} 
      b_{i,\alpha}^{\dagger}b_{j,\alpha} + 
      U_{\rm{eff}}^{} \sum_{i,\alpha \neq \beta}n_{i,\alpha}n_{i,\beta} 
  \nonumber \\
  &+& V_{\rm{eff}}^{} \sum_{\langle i,j \rangle,\alpha \neq \beta}
      n_{i,\alpha}n_{j,\beta},
\label{hami_B}
\end{eqnarray}
where $b_{i,\alpha}$ annihilates a bosonic fermion pair on orbital $\alpha$ in the $i$th lattice site, $t_{\rm{eff}}^{}$ ($=-2t^{2}/U$) represents the effective hopping integral of the boson, and $n_{i,\alpha}=b_{i,\alpha}^{\dagger}b_{i,\alpha}$. 
The effective interactions between two bosonic fermion pairs in the same lattice site and in the neighboring sites are denoted as $U_{\rm{eff}}^{}$ ($=-2U'+J-D$) and $V_{\rm{eff}}^{}=-2t_{\rm{eff}}^{2}/U_{\rm{eff}}$, respectively. 
Under the condition $U'=J=U/2(<0)$, we obtain $U_{\rm{eff}}^{}>0$ and thus $V_{\rm{eff}}^{}<0$. The repulsive $U_{\rm{eff}}^{}$ and attractive $V_{\rm{eff}}^{}$ prefer the ordered state where one bosonic fermion pair occupies each site with two orbitals being occupied alternately in the neighboring sites: the orbital AF state of the boson. 
Note that the occupied (unoccupied) state of the boson in the higher orbital represents the empty (doubly-occupied) state of the fermionic atoms in the original Hamiltonian, because of a particle-hole transformation. 
Accordingly, the orbital AF state of boson represents that the two orbitals in the same site are occupied by two pairs of fermionic atoms with different pseudospins and the neighboring site is empty in a viewpoint of the original Hamiltonian. This state is nothing but the ADW state. We have thus demonstrated that the effective boson model yields the ADW state in agreement with the numerical results.

\begin{figure}[htb]
\begin{center}
\includegraphics[scale=0.65]{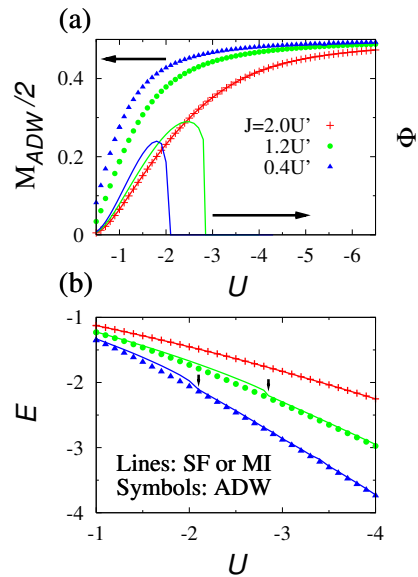}
\caption{(Color Online) (a) The ADW order parameter $M_{\rm ADW}$ and the superfluid order parameter $\Phi$ as functions of the attractive interaction $U$ for several values of $J$ in $U'=U/2$ and $D=0$. For $J=2U'$, $M_{\rm{ADW}}/2$ and $\Phi$ show the same $U$ dependence. 
(b) The energies of the ADW, superfluid, and Mott insulating states are compared for several values of $J$ in $U'=U/2$ and $D=0$. At the arrows, the superfluid (SF) states change into the Mott-insulating (MI) states. For $J=2U'$, the energies of the superfluid and ADW states are degenerate. }
\label{eff}
\end{center}
\end{figure}

The effective Hamiltonian \eqref{hami_B} further suggests that for $J=2U'$ and $D=0$ the effective repulsion vanishes and $V_{\rm{eff}}^{B}$ is strongly enhanced. At these parameters, the superfluid state and/or the ADW state are expected to appear. 
To see the ordered state when the system approaches the parameters $J=2U'$ and $D=0$, we calculate the order parameters of the superfluid state $\Phi$ and the ADW state $M_{\rm{ADW}}$, releasing $J$ from the restriction $U'=J=U/2$. In Fig. \ref{eff}, the results for several values of $J$ in $U'=U/2$ for $D=0$ are shown as functions of $U$. 
We find that $\Phi$ keeps nonzero values only for $J=2U'$. In this case, $M_{\rm{ADW}}/2$ and $\Phi$ exhibit the same $U$ dependence as shown in Fig. \ref{eff}(a). Furthermore, the energies of the superfluid and ADW states are degenerate at $J=2U'$ and $D=0$, although the energy of the ADW state is the lowest for other parameters as shown in Fig. \ref{eff}(b). These results indicate that both states coexist: a supersolid state emerges. 
For the system with $U'=J=U/2$, the coexisting region of the superfluid and ADW states lies along the boundary of both phases shown in Fig. \ref{PD}. 

In real systems, there is a confinement potential. Recently, the effects of a harmonic confinement potential were investigated for the single-band attractive Hubbard model on the square lattice \cite{Koga2008}. It was shown that a harmonic confinement potential plays an essential role in stabilizing the supersolid state and that the doughnuts-like region of the supersolid state emerges between the ADW and superfluid phases. 
In the present system, the ADW and band-insulating states appear in addition to the superfluid state owing to the two-band effects. This is in contrast with the phases of the single-band attractive Hubbard model without a confinement potential, where only the superfluid state appears at half-filling as mentioned before. 
For the two-band attractive lattice fermion systems with a confinement potential, a richer phase diagram including a supersolid state is expected to appear. 
The investigation of this issue with large numerical calculations is our next study. 

\section{SUMMARY}
We have investigated the ordered state of ultracold fermionic atoms in three-dimensional optical lattices at half-filling using two-site dynamical mean-field theory. Because of the two-orbital effects, the ADW state emerges for the attractive interaction region, while the AF state emerges for the repulsive interaction region. 
We have shown that the ADW and AF states are more stable than the Mott insulating states. However, the ADW and AF states may become unstable by thermal fluctuations. Actually, it was shown for the single-band Hubbard model that the transition from the  AF state to the Mott insulator occurs around $T/T_{\rm F} \sim O(10^{-1})$ at half-filling with increasing temperature \cite{Georges1996}. Here $T_{\rm F}$ is the Fermi temperature. 
The experiments for cold fermionic atoms in optical lattices have been performed in temperatures $T/T_{\rm F} \sim O(10^{-1})$ \cite{Kohl2005,Jordens2008}. 
Accordingly, the competition between the Mott insulating state and the ordered state (the ADW state or the AF state) at finite temperatures is expected. Detailed investigations for the ordered state of multiband fermionic systems at finite temperatures are our future work.

\section{ACKNOWLEDGEMENTS}
We thank A. Koga, A. Yamamoto, and M. Yamashita for useful comments and valuable discussions. 
Numerical computations were carried out at the Supercomputer Center, the Institute for Solid State Physics, University of Tokyo. 
K.I. was supported by the Japan Society for the Promotion of Science. This work was supported by a Grant-in-Aid (No. 20540390) for Scientific Research from the Ministry of Education, Culture, Sports, Science and Technology, Japan.

\bibliography{apssamp}

\begin{thebibliography}{00}

\bibitem{Chen2005} For a review, see Q. Chen, J. Stajic, S. Tan, and K. Levin,  Phys. Rep. {\bf 412}, 1 (2005).

\bibitem{Kohl2005} M. K\"{o}hl, H. Moritz, T. St\"{o}ferle, K. G\"{u}nter, and T. Esslinger,  Phys. Rev. Lett. {\bf 94}, 080403 (2005).

\bibitem{Jordens2008} R. J\"{o}rdens, N. Strohmaier, K. G\"{u}nter, H. Moritz,  and T. Esslinger,  Nature {\bf 455}, 204 (2008); U. Schneider, L. Hackerm\"{u}ller, S. Will, Th. Best, I. Bloch, T. A. Costi, R. W. Helmes, D. Rasch, and A. Rosch, Science {\bf 322}, 1520 (2008)

\bibitem{Chin2006} J. K. Chin, D. E. Miller, Y. Liu, C. Stan, W. Setiawan, C. Sanner, K. Xu, and W. Ketterle,  Nature {\bf 443}, 961 (2006).

\bibitem{Kollath2006} C. Kollath, A. Iucci, I. McCulloch, and T. Giamarchi, Phys. Rev. A {\bf 74}, 041604(R) (2006).

\bibitem{Zhai2007} H. Zhai and T. L. Ho, Phys. Rev. Lett. {\bf 99}, 100402 (2007).

\bibitem{Moon2007} E. G. Moon, P. Nikolic, and S. Sachdev, Phys. Rev. Lett. {\bf 99}, 230403 (2007).

\bibitem{Chien2008} C. C. Chien, Y. He, Q. Chen, and K. Levin, Phys. Rev. A {\bf 77}, 011601(R) (2008). 

\bibitem{Helmes2008} R. W. Helmes, T. A. Costi, and A. Rosch, Phys. Rev. Lett. {\bf 100}, 056403 (2008).

\bibitem{Higashi} T. Higashiyama, K. Inaba, and S. Suga,  Phys. Rev. A {\bf 77}, 043624 (2008).

\bibitem{Goldman} N. Goldman, Phys. Rev. A {\bf 77}, 053406 (2008).

\bibitem{Andersen} B. M. Andersen and G. M. Bruun, Phys. Rev. A {\bf 76}, 041602(R) (2007).

\bibitem{Wu} K. Wu and H. Zhai, Phys. Rev. B {\bf 77}, 174431 (2008).

\bibitem{Koga2008} A. Koga, T. Higashiyma, K. Inaba, S. Suga, and N. Kawakami,  J. Phys. Soc. Jpn. {\bf 77}, 073602 (2008); Phys. Rev. A {\bf 78}, (2009) in press.

\bibitem{Snoek2008} M. Snoek, I. Titvinidze, C. T\"{o}ke, K. Byczuk, and W. Hofstetter, New. J. Phys. {\bf 10}, 093008 (2008).

\bibitem{Georges1996} A. Georges, G. Kotliar, W. Krauth, and M. J. Rozenberg,  Rev. Mod. Phys. {\bf 68}, 13 (1996).

\bibitem{Zwerger2003} W. Zwerger,  J. Opt. B {\bf 5}, S9 (2003).

\bibitem{Hofstetter2006} W. Hofstetter,  Philos. Mag. {\bf 86}, 1891 (2006).

\bibitem{Potthoff2001} M. Potthoff,  Phys. Rev. B {\bf 64}, 165114 (2001).

\bibitem{Ono2003} Y. Ono, M. Potthoff, and R. Bulla,  Phys. Rev. B {\bf 67}, 035119 (2003).

\bibitem{Koga2004}
A. Koga, N. Kawakami, T. M. Rice, and M. Sigrist,  Phys. Rev. Lett. {\bf 92}, 216402 (2004).

\bibitem{Chitra}
R. Chitra and G. Kotliar,  Phys. Rev. Lett. {\bf 83}, 2386 (1999).

\bibitem{Garg2005} A. Garg, H. R. Krishnamurthy, and M. Randeria,  Phys. Rev. B {\bf 72}, 024517 (2005).

\bibitem{Scalettar1989} R. T. Scalettar, E. Y. Loh, J. E. Gubernatis, A. Moreo, S. R. White, D. J. Scalapino, R. L. Sugar, and E. Dagotto,  Phys. Rev. Lett. {\bf 62}, 1407 (1989).

\bibitem{Capone2002} M. Capone, C. Castellani, and M. Grilli,  Phys. Rev. Lett. {\bf 88}, 126403 (2002).



\bibitem{Toschi2005} A. Toschi, M. Capone, and C. Castellani,  Phys. Rev. B {\bf 72}, 235118 (2005).

\bibitem{Kyung2006} B. Kyung, A. Georges, and A. M. S. Tremblay,  Phys. Rev. B {\bf 74}, 024501 (2006).

\bibitem{Koga2002} A. Koga, Y. Imai, and N. Kawakami, Phys. Rev. B {\bf 66}, 165107 (2002).

\bibitem{Werner2007} P. Werner and A. J. Millis, Phys. Rev. Lett. {\bf 99}, 126405 (2007).



\end{thebibliography}


\end{document}